\documentclass[showpacs,twocolumn,floatfix]{revtex4}
\usepackage{graphicx}
\usepackage{amsmath}
\begin{document}
\title{Production and detection of entangled electron-hole pairs in a
degenerate electron gas}
\author{C. W. J. Beenakker, C. Emary, M. Kindermann, and J. L. van Velsen}
\affiliation{Instituut-Lorentz, Universiteit Leiden, P.O. Box 9506, 2300 RA
Leiden, The Netherlands}
\date{6 May 2003}
\begin{abstract}
We demonstrate theoretically that the shot noise produced by a tunnel barrier
in a two-channel conductor violates a Bell inequality. The non-locality is
shown to originate from entangled electron-hole pairs created by tunneling
events --- without requiring electron-electron interactions. The degree of
entanglement (concurrence) equals $2(T_{1}T_{2})^{1/2}(T_{1}+T_{2})^{-1}$, with
$T_{1},T_{2}\ll 1$ the transmission eigenvalues. A pair of edge channels in the
quantum Hall effect is proposed as experimental realization.
\end{abstract}
\maketitle

The controlled production and detection of entangled particles is the first
step on the road towards quantum information processing \cite{Ter03}. In optics
this step was taken long ago \cite{Asp81}, but in the solid state it remains an
experimental challenge. A variety of methods to entangle electrons have been
proposed, based on quite different physical mechanisms \cite{Egu02}. A common
starting point is a spin-singlet electron pair produced by interactions, such
as the Coulomb interaction in a quantum dot \cite{Bur00,Oli02,Sar02}, the
pairing interaction in a superconductor \cite{Les01,Rec01,Ben02}, or Kondo
scattering by a magnetic impurity \cite{Cos01}. A very recent proposal based on
orbital entanglement also makes use of the superconducting pairing interaction
\cite{Sam03}.

It is known that photons can be entangled by means of linear optics using a
beam splitter \cite{Kni01,Sch01,Kim02}. The electronic analogue would be an
entangler that is based entirely on single-electron physics, without requiring
interactions. But a direct analogy with optics fails: Electron reservoirs are
in local thermal equilibrium, while in optics a beam splitter is incapable of
entangling photons from a thermal source \cite{Xia02}. That is why previous
proposals \cite{Cos01,Bos02} to entangle electrons by means of a beam splitter
start from a two-electron Fock state, rather than a many-electron thermal
state. To control the extraction of a single pair of electrons from an electron
reservoir requires strong Coulomb interaction in a tightly confined area, such
as a semiconductor quantum dot or carbon nanotube \cite{Egu02}. Indeed, it has
been argued \cite{Cht02} that one can not entangle a spatially separated
current of electrons from a normal (not-superconducting) source without
recourse to interactions.

What we would like to propose here is an altogether different, interaction-free
source of entangled quasiparticles in the solid state. The entanglement is not
between electron pairs but between electron-hole pairs in a degenerate electron
gas. The entanglement and spatial separation are realized purely by elastic
scattering at a tunnel barrier in a two-channel conductor. We quantify the
degree of entanglement by calculating how much the current fluctuations violate
a Bell inequality.

\begin{figure}
\includegraphics[width=8cm]{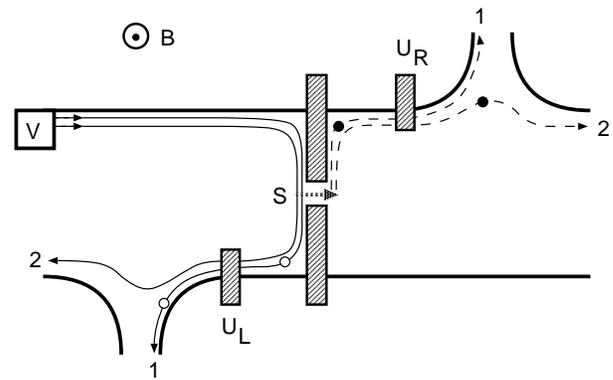}
\caption{
Schematic description of the method to produce and detect entangled edge
channels in the quantum Hall effect. The thick black lines indicate the
boundaries of a two-dimensional electron gas. A strong perpendicular magnetic
field $B$ ensures that the transport near the Fermi level $E_{F}$ takes place
in two edge channels, extended along a pair of equipotentials (thin solid and
dashed lines, with arrows that give the direction of propagation). A split gate
electrode (dashed rectangles at the center) divides the conductor into two
halves, coupled by tunneling through a narrow opening (dashed arrow, scattering
matrix $S$). If a voltage $V$ is applied between the two halves, then there is
a narrow energy range $eV$ above $E_{F}$ in which the edge channels are
predominantly filled in the left half (solid lines) and predominantly empty in
the right half (dashed lines). Tunneling events introduce filled states in the
right half (black dots) and empty states in the left half (open circles). The
entanglement of these particle-hole excitations is detected by the violation of
a Bell inequality. This requires two gate electrodes to locally mix the edge
channels (scattering matrices $U_{L}$, $U_{R}$) and two pair of contacts $1,2$
to separately measure the current in each transmitted and reflected edge
channel.
\label{bell_circuit}
}
\end{figure}

Any two-channel conductor containing a tunnel barrier could be used in
principle for our purpose, and the analysis which follows applies generally.
The particular implementation described in Fig.\ \ref{bell_circuit} uses edge
channel transport in the integer quantum Hall effect \cite{Bee91}. It has the
advantage that the individual building blocks have already been realized
experimentally for different purposes. If the two edge channels lie in the same
Landau level, then the entanglement is between the spin degrees of freedom.
Alternatively, if the spin degeneracy is not resolved by the Zeeman energy and
the two edge channels lie in different Landau levels, then the entanglement is
between the orbital degrees of freedom. The beam splitter is formed by a split
gate electrode, as in Ref.\ \cite{Hen99}. In Fig.\ \ref{bell_circuit} we show
the case that the beam splitter is weakly transmitting and strongly reflecting,
but it could also be the other way around. To analyze the Bell inequality an
extra pair of gates mixes the orbital degrees of freedom of the outgoing states
independently of the incoming states. (Alternatively, one could apply a local
inhomogeneity in the magnetic field to mix the spin degrees of freedom.)
Finally, the current in each edge channel can be measured separately by using
their spatial separation, as in Ref.\ \cite{Wee89}. (Alternatively, one could
use the ferromagnetic method to measure spin current described in Ref.\
\cite{Egu02}.)

Electrons are incident on the beam splitter from the left in a range $eV$ above
the Fermi energy $E_{F}$. (The states below $E_{F}$ are all occupied at low
temperatures, so they do not contribute to transport properties.) The incident
state has the form
\begin{equation}
|\Psi_{\rm in}\rangle=\prod_{0<\varepsilon<eV}a_{{\rm
in},1}^{\dagger}(\varepsilon)a_{{\rm in},2}^{\dagger}(\varepsilon)|0\rangle.
\label{Psiindef}
\end{equation}
The fermion creation operator $a^{\dagger}_{{\rm in},i}(\varepsilon)$ excites
the $i$-th channel incident from the left at energy $\varepsilon$ above the
Fermi level. Similarly, $b^{\dagger}_{{\rm in},i}(\varepsilon)$ excites a
channel incident from the right. Each excitation is normalized such that it
carries unit current. It is convenient to collect the creation operators in two
vectors $a^{\dagger}_{\rm in}$, $b^{\dagger}_{\rm in}$ and to use a matrix
notation,
\begin{equation}
|\Psi_{\rm in}\rangle=
\prod_{\varepsilon}
\left(\begin{array}{l}
a_{\rm in}^{\dagger}\\b_{\rm in}^{\dagger}
\end{array}\right)
\left(\begin{array}{cc}
\frac{1}{2}\sigma_{y}&0\\0&0
\end{array}\right)
\left(\begin{array}{l}
a_{\rm in}^{\dagger}\\b_{\rm in}^{\dagger}
\end{array}\right)|0\rangle,
\label{Psiinmatrix}
\end{equation}
with $\sigma_{y}$ a Pauli  matrix.

The input-output relation of the beam splitter is
\begin{equation}
\left(\begin{array}{l}
a_{\rm out}\\b_{\rm out}
\end{array}\right)
=
\left(\begin{array}{cc}
r&t'\\t&r'
\end{array}\right)
\left(\begin{array}{l}
a_{\rm in}\\b_{\rm in}
\end{array}\right). \label{inputoutput}
\end{equation}
The $4\times 4$ unitary scattering matrix $S$ has $2\times 2$ submatrices
$r,r',t,t'$ that describe reflection and transmission of states incident from
the left or from the right. Substitution of Eq.\ (\ref{inputoutput}) into Eq.\
(\ref{Psiinmatrix}) gives the outgoing state
\begin{eqnarray}
|\Psi_{\rm out}\rangle&=&\prod_{\varepsilon}\bigl(a_{{\rm out}}^{\dagger}
r\sigma_{y} t^{\rm T}b_{{\rm out}}^{\dagger}
+[r\sigma_{y} r^{\rm T}]_{12}^{\vphantom\dagger} a_{{\rm
out},1}^{\dagger}a_{{\rm out},2}^{\dagger}
\nonumber\\
&&\mbox{}+[t\sigma_{y} t^{\rm T}]_{12}^{\vphantom\dagger} b_{{\rm
out},1}^{\dagger}b_{{\rm out},2}^{\dagger}\bigr)|0\rangle. \label{Psioutdef}
\end{eqnarray}
The superscript ``T'' indicates the transpose of a matrix.

To identify the entangled electron-hole excitations we transform from particle
to hole operators at the left of the beam splitter: $c_{{\rm out},i}=a_{{\rm
out},i}^{\dagger}$. The new vacuum state is $a_{{\rm out},1}^{\dagger}a_{{\rm
out},2}^{\dagger}|0\rangle$. To leading order in the transmission matrix the
outgoing state becomes
\begin{eqnarray}
&&|\Psi_{\rm
out}\rangle=\prod_{\varepsilon}\bigl(\sqrt{w}|\Phi\rangle+\sqrt{1-w}|0\rangle\bigr),
\label{wdef}\\
&&|\Phi\rangle=w^{-1/2}c_{\rm out}^{\dagger}\gamma b_{\rm
out}^{\dagger}|0\rangle,\;\;
\gamma=\sigma_{y} r\sigma_{y}t^{\rm T}. \label{gammadef}
\end{eqnarray}
It represents a superposition of the vacuum state and a particle-hole state
$\Phi$ with weight $w={\rm Tr}\,\gamma\gamma^{\dagger}$.

The degree of entanglement of $\Phi$ is quantified by the concurrence
\cite{Woo98,Note3},
\begin{equation}
{\cal C}=2\sqrt{{\rm Det}\,\gamma\gamma^{\dagger}}/{\rm
Tr}\,\gamma\gamma^{\dagger},\label{Cdef}
\end{equation}
which ranges from $0$ (no entanglement) to $1$ (maximal entanglement).
Substituting Eq.\ (\ref{gammadef}) and using the unitarity of the scattering
matrix we find after some algebra that
\begin{eqnarray}
{\cal
C}&=&\frac{2\sqrt{(1-T_{1})(1-T_{2})T_{1}T_{2}}}{T_{1}+T_{2}-2T_{1}T_{2}}
\nonumber\\
&\approx&2\sqrt{T_{1}T_{2}}/(T_{1}+T_{2})\;\;{\rm if}\;\;T_{1},T_{2}\ll
1.\label{Cresult}
\end{eqnarray}
The concurrence is entirely determined by the eigenvalues $T_{1},T_{2}\in(0,1)$
of the transmission matrix product $t^{\dagger}t=\openone-r^{\dagger}r$. The
eigenvectors do not contribute. Maximal entanglement is achieved if the two
transmission eigenvalues are equal: ${\cal C}=1$ if $T_{1}=T_{2}$.

The particle-hole entanglement is a nonlocal correlation that can be detected
through the violation of a Bell inequality. We follow the formulation in terms
of irreducible current correlators of Samuelsson, Sukhorukov, and B\"{u}ttiker
\cite{Sam03}. In the tunneling regime considered here that formulation is
equivalent to the original formulation in terms of coincidence counting rates
\cite{Bel64}. The tunneling assumption is essential: If $T_{1},T_{2}$ are not
$\ll 1$ one can not violate the Bell inequality without coincidence detection
\cite{Cht02}.

The quantity $C_{ij}=\int_{-\infty}^{\infty}dt\,\overline{\delta
I_{L,i}(t)\delta I_{R,j}(0)}$ correlates the time-dependent current
fluctuations $\delta I_{L,i}$ in channel $i=1,2$ at the left with the current
fluctuations $\delta I_{R,j}$ in channel $j=1,2$ at the right. It can be
measured directly in the frequency domain as the covariance of the
low-frequency component of the current fluctuations. At low temperatures
($kT\ll eV$) the correlator has the general expression \cite{But90}
\begin{equation}
C_{ij}=-(e^{3}V/h)|(rt^{\dagger})_{ij}|^{2}.\label{Cijdef}
\end{equation}
We need the following rational function of correlators:
\begin{equation}
E=\frac{C_{11}+C_{22}-C_{12}-C_{21}}{C_{11}+C_{22}+C_{12}+C_{21}}=\frac{{\rm
Tr}\,\sigma_{z}rt^{\dagger}\sigma_{z}tr^{\dagger}}{{\rm
Tr}\,r^{\dagger}rt^{\dagger}t}.\label{Edef}
\end{equation}
By mixing the channels locally in the left and right arm of the beam splitter,
the transmission and reflection matrices are transformed as $r\rightarrow
U_{L}r$, $t\rightarrow U_{R}t$, with unitary $2\times 2$ matrices
$U_{L},U_{R}$. The correlator transforms as
\begin{equation}
E(U_{L},U_{R})=\frac{{\rm
Tr}\,U_{L}^{\dagger}\sigma_{z}U_{L}^{\vphantom\dagger}rt^{\dagger}U_{R}^{\dagger}
\sigma_{z}U_{R}^{\vphantom\dagger}tr^{\dagger}}{{\rm Tr}\,r^{\dagger}rt^{\dagger}t}.
\label{EULUR}
\end{equation}
The Bell-CHSH (Clauser-Horne-Shimony-Holt) parameter is \cite{Bel64,Note2}
\begin{equation}
{\cal E}=
E(U_{L},U_{R})+E(U'_{L},U_{R})+E(U_{L},U'_{R})-E(U'_{L},U'_{R}).\label{CHSHdef}
\end{equation}
The state is entangled if $|{\cal E}|>2$ for some set of unitary matrices
$U_{L}, U_{R}, U'_{L}, U'_{R}$. By repeating the calculation of Ref.\
\cite{Pop92} we find the maximum \cite{Note1}
\begin{equation}
{\cal E}_{\rm
max}=2[1+4T_{1}T_{2}(T_{1}+T_{2})^{-2}]^{1/2}>2.\label{Emaxresult}
\end{equation}
Comparison with Eq.\ (\ref{Cresult}) confirms the expected relation ${\cal
E}_{\rm max}=2(1+{\cal C}^{2})^{1/2}$ between the concurrence and the maximal
violation of the CHSH inequality \cite{Gis91}.

As a final consistency check we consider the effect of dephasing \cite{Note4}.
Dephasing is modeled by introducing random phase factors in each edge channel,
which amounts to the substitutions
\begin{equation}
U_{L}\rightarrow U_{L}
\left(\begin{array}{cc}
e^{i\phi_{1}}&0\\
0&e^{i\phi_{2}}
\end{array}\right),\;\;
U_{R}\rightarrow U_{R}
\left(\begin{array}{cc}
e^{i\psi_{1}}&0\\
0&e^{i\psi_{2}}
\end{array}\right).
\label{phasefactors}
\end{equation}
We average $E(U_{L},U_{R})$ over the random phases, uniformly in $(0,2\pi)$,
and find
\begin{equation}
{\cal E}_{\rm max}=\frac{2|{\rm
Tr}\,\sigma_{z}rt^{\dagger}\sigma_{z}tr^{\dagger}|}{{\rm
Tr}\,r^{\dagger}rt^{\dagger}t}\leq 2.\label{Emaxaverage}
\end{equation}
So for strong dephasing there is no violation of the Bell inequality $|{\cal
E}|\leq 2$, which is as it should be.

In conclusion, we have demonstrated theoretically that a tunnel barrier creates
spatially separated currents of entangled electron-hole pairs in a degenerate
electron gas. Because no Coulomb or pairing interaction is involved, this is an
attractive alternative to existing proposals for the interaction-mediated
production of entanglement in the solid state. We have described a possible
realization using edge channel transport in the quantum Hall effect, which
makes use of existing technology. There is a remarkable contrast with quantum
optics, where a beam splitter can not create entanglement if the source is in
local thermal equilibrium. This might well explain why the elementary mechanism
for entanglement production described here was not noticed before.

We have benefitted from correspondence with P. Samuelsson. This work was
supported by the Dutch Science Foundation NWO/FOM and by the U.S. Army Research
Office (Grant No. DAAD 19-02-0086).

\end{document}